\documentclass{aa}
\usepackage{graphicx}
\usepackage{longtable}
\usepackage{lscape}

\def\gsim{\;\lower.6ex\hbox{$\sim$}\kern-7.75pt\raise.65ex\hbox{$>$}\;}
\def\lsim{\;\lower.6ex\hbox{$\sim$}\kern-7.75pt\raise.65ex\hbox{$<$}\;}

\begin{document}
\title{Na-O Anticorrelation and HB. IV. Detection of He-rich and He-poor
stellar populations in the globular cluster NGC 6218
\thanks{
Based on observations collected at ESO telescopes under programme 073.D-0211. 
Tables 2, 3, 5 are only available in electronic form
at the CDS via anonymous ftp to cdsarc.u-strasbg.fr (130.79.128.5)
or via http://cdsweb.u-strasbg.fr/cgi-bin/qcat?J/A+A/}
 }

\author{
E. Carretta\inst{1},
A. Bragaglia\inst{1},
R.G. Gratton\inst{2},
G. Catanzaro\inst{3},
F. Leone\inst{3,4},
E. Sabbi\inst{5},
S. Cassisi\inst{6},
R. Claudi\inst{2},
F. D'Antona\inst{7},
P. Fran\c cois\inst{8},
G. James\inst{9},
\and
G. Piotto\inst{10} 
}

\authorrunning{E. Carretta et al.}
\titlerunning{Na-O anticorrelation in NGC 6218}

\offprints{E. Carretta, eugenio.carretta@oabo.inaf.it}

\institute{
INAF-Osservatorio Astronomico di Bologna, Via Ranzani 1, I-40127
 Bologna, Italy
\and
INAF-Osservatorio Astronomico di Padova, Vicolo dell'Osservatorio 5, I-35122
 Padova, Italy
\and
INAF-Osservatorio Astrofisico di Catania, Via S.Sofia 78, I-95123 Catania, Italy
\and
Dipartimento di Fisica ed Astronomia, Universit\`a di Catania, Via S.Sofia 78, 
 I-95123 Catania, Italy
\and
Space Telescope Science Institute, 3700 San Martin Drive,
Baltimore, MD 21218, USA
\and
INAF-Osservatorio Astronomico di Collurania, Via M. Maggini, I-64100 Teramo,
Italy
\and
INAF-Osservatorio Astronomico di Roma, Via Frascati 33, I-00040 Monteporzio,
Italy
\and
Observatoire de Paris, 61 Avenue de l'Observatoire, F-75014 Paris, France
\and
European Southern Observatory, Alonso de Cordova 3107, Vitacura, Santiago, Chile
\and
Dipartimento di Astronomia, Universit\`a di Padova, Vicolo dell'Osservatorio 2,
I-35122 Padova, Italy
  }

\date{8 jan 2007}

\abstract{
We used the multifiber spectrograph FLAMES on the ESO Very Large Telescope UT2
to derive atmospheric parameters, metallicities and abundances of O
and Na for 79 red giant stars in the Galactic globular cluster NGC~6218
(M~12). We analyzed stars in the magnitude range from about 1 mag below the
bump to the tip of the Red Giant Branch. 
The average metallicity we derive is [Fe/H]=$-1.31\pm 0.004\pm 0.028$ dex 
(random and systematic errors, respectively), with a
very small star-to-star scatter ($rms=0.033$ dex), from moderately
high-resolution Giraffe spectra. This is the first extensive spectroscopic 
abundance analysis in this cluster. Our results indicate that NGC 6218 is very
homogeneous as far as heavy elements are concerned.
On the other hand, light elements involved in the well known proton-capture 
reactions of H-burning at high temperature, such as O and Na, show large 
variations, anticorrelated with each other, at all luminosities along the red 
giant branch. The conclusion is that the Na-O anticorrelation must be
established in early times at the cluster formation. 
We interpret the variation of Na found near the RGB-bump as the effect of two
distinct populations having different bump luminosities, as predicted for
different He content. To our knowledge, NGC~6218 is the first GC where such a
signature has been spectroscopically detected, when combined with consistent
and homogeneous data obtained for NGC~6752 to gain in statistical significance.
} 

\maketitle

\section{Introduction}

This is the fourth paper of a series aimed at studying the mechanisms of
formation and early evolution of Galactic globular clusters (GCs). To reach
this goal we are trying to uncover and establish the possible existence of a 
second generation of stars in these old stellar aggregates.
The motivations and the general strategy have already
been explained in the first papers of the series, particularly in Carretta
et al. (2006a, hereafter Paper I) dedicated to NGC 2808.

It has been recently realized that a few well known abundance "anomalies", one
for all the anticorrelation between abundances of Na and O, are present
also among unevolved or scarcely evolved cluster stars (Gratton et al. 2001, 
Ramirez \& Cohen 2003, Cohen \& Melendez 2005, Carretta et al. 2005).
The chemical composition of these stars shows that the alterations observed in
the abundances of the light elements C, N, O, Al, Mg cannot be attributed to an
extensive phenomenon of evolutionary extra-mixing, since these stars
lack the possibility to mix internal nucleosynthetic products to the surface
(for reviews see Kraft 1994 and more recently Gratton et al. 2004).
It naturally follows that an origin from stars other than the ones we
are observing has to be accepted (the so called primordial origin).
The best candidates are an earlier generation of intermediate-mass  
Asymptotic Giant Branch (AGB) stars, able to pollute the material from which
the currently observed stars were born from. 
We then expect to be able to see  differences in the abundances for
those elements involved in nuclear burning in those pristine stars, i.e. those
involved in high temperature proton-capture reactions in the complete 
CNO cycle (Ne-Na and Mg-Al chains).

It is interesting to note that these anomalies are a signature strictly
confined to the high density environment typical of globular clusters: field
stars are known to show only the modifications in C and N abundances as
predicted by the two typical mixing episodes (first dredge-up and a second
mixing after the bump on the RGB) expected for Population II low mass stars,
and no alterations in the abundance of heavier nuclei such as O, Na (Charbonnel
1994, Gratton et al. 2000). This striking difference supports the idea that this
kind of anomalies might be intimately related to the very same formation of a
globular cluster (see e.g. Carretta 2006).
On the other hand, if this conclusion is true, we might expect that other
physical properties of GCs may show a connection with their chemical signature.

A first surprising discovery was made by  Carretta (2006): using more limited
samples than those we are collecting in our present survey, he pointed out that
a quite good correlation does exist between the spread in the abundance of
proton-capture elements and orbital parameters of GCs. In particular, chemical
anomalies are found to be more extended in clusters having large-sized orbits
and longer periods and in those with larger inclination angles of the orbit
with respect to the Galactic plane.

In our study, we 
are  performing a systematic analysis of a large number of stars
(about 100 per cluster) in about 20 GCs, determining accurately and
homogeneously abundances of Na and O. We have chosen clusters of
different horizontal branch (HB) morphologies, metallicities, 
concentration, densities, et cetera, to understand
whether there is some connection between the presence and extent of the
anticorrelations and some properties of the globular clusters. 

To reduce concerns related to model atmospheres, we are not observing stars
near the red giant branch (RGB) tip. To sample a CMD region 
populated enough and reach the required number of objects, in some clusters we
have to select stars in a rather extended range of magnitudes, sampling the 
cluster
giants both above and below the so-called RGB-bump, i.e. the bump in the  
luminosity function of RGB stars.
NGC~6218 is one of these cases and we were able to study in details the region
near the RGB-bump, where e.g., effects due to different He contents are enhanced and
observable thanks to the slowing of  stellar evolutionary rates. This allows us
to uncover for the first time the spectroscopic evidence of distinct stellar
populations with likely different He content.

In the framework of the present project we have already examined the youngish 
and peculiar cluster NGC~2808 (Paper I),
the more "normal", intermediate-metallicity, blue HB cluster NGC~6752 
(Carretta et al. 2006, hereafter Paper II), and the anomalous bulge cluster 
NGC~6441 (Gratton et al. 2006, Paper III).  
Here we present our results on the Na-O anticorrelation among
RGB stars in NGC~6218 (M~12), studied for the first time in this cluster.  

From the Harris (1996) catalogue NGC~6218 has an absolute visual magnitude
$M_V=-7.32$, a distance from the Sun of 4.9 kpc and from the Galactic centre of
4.5 kpc, and a totally blue HB  [$(B-R)/(B+V+R)=0.97$]
without any RR Lyrae star. Its age seems in line with the ones of the oldest
Galactic GCs (Rosenberg et al. 1999)
The only known peculiarity of NGC~6218 is its very flat mass function (De
Marchi et al. 2006), indicative of a very efficient mass segregation
and tidal stripping.

Up to now, NGC~6218 had not been the subject of many detailed spectroscopic 
works, not even to determine its metallicity 
\footnote{We adopt the  usual spectroscopic notation, $i.e.$ 
[X]= log(X)$_{\rm star} -$ log(X)$_\odot$ for any abundance quantity X, and 
log $\epsilon$(X) = log (N$_{\rm X}$/N$_{\rm H}$) + 12.0 for absolute number
density abundances.}.
Before the present work\footnote{
After the first version of the present paper was submitted, Jonhson \&
Pilachowski (2006) presented the analysis of 21 RGB and AGB stars, observed with
Hydra on the 3.5m WIYN telescope at resolution $R\sim15000$. They derived and
average metallicity [Fe/H]=$-1.54$ ($\sigma=0.09$) dex and measured abundances
of several elements ($\alpha$-, Fe-peak, n-capture elements); in particular, they
obtained abundances of Na, but not of O. Since they already presented a quite
detailed comparison of their results with ours, we will not repeat it here.},
only three stars in this cluster had their abundances
derived from high resolution spectra. Klochkova \& Samus (2001) observed the
post-AGB star K~413 (ZNG~8), determining [Fe/H]$=-1.38$, [Na/Fe]$=+0.44$ and a
very large overabundance of O ([O/Fe]$\simeq+2.2$).  Mishenina et al. (2003)
analyzed star K~118 (ZNG~4), finding
[Fe/H]=$-1.36$;  they did not measure Na or O. Jasniewicz et al. (2004)
presented the analysis of one UV-bright star (ZNG~7, or K~329) in this
cluster, a probable AGB star which shows an underabundance of Fe
([Fe/H]$=-1.75$) and several other peculiarities; neither Na or O  abundances
were determined.

Other literature values for the cluster's metallicity include the one by Brodie
\& Hanes (1986) based on integrated light and the use of indices which
resulted in [Fe/H]$=-1.21\pm0.02$, and the one by Rutledge et al. (1997) based
on the Ca IR triplet, which is [Fe/H]$=-1.14\pm0.05$ (when the method is
calibrated on the Carretta \& Gratton 1997 metallicity scale) or
[Fe/H]$=-1.40\pm0.07$ (when calibrated on the Zinn \& West 1984 scale). 

The present paper is organized as follows: an outline of the observations is
given in the next Section; the derivation of atmospheric  parameters and the
analysis are described in Sect. 3, whereas error estimates and cosmic scatter
in metallicity are discussed  in
Sect. 4.  Sect. 5 is devoted to the results for the Na-O anticorrelation.
In Sect. 6 we present the evidences for the existence of two distinct
population of He-rich and He-poor stars along the RGB of NGC~6218. 
The relation between chemistry and orbital parameters of the cluster is shown
in Sect. 7. Finally, a summary is presented in Sect. 8.

\section{Observations, target selection and membership}

Our data were collected in Service mode 
with the ESO high resolution multifiber spectrograph 
FLAMES/GIRAFFE (Pasquini et al. 2002) mounted on the VLT. Observations were done
with two GIRAFFE setups, the high-resolution gratings HR11 (centered at
5728~\AA)  and HR13 (centered at 6273~\AA) to measure the Na doublets at
5682-5688~\AA\ and 6154-6160~\AA\ and the [O {\sc i}] forbidden lines at 6300, 
6363~\AA, respectively. The resolution is R=24200 (for HR11) and R=22500  (for
HR13), at the center of the spectra. We used one single exposure of 2700 seconds
for each grating.

\begin{table}
\caption{Log of the observations for NGC 6218. Date and time are UT, exposure
times are in seconds. For both exposures the field center is at
RA(2000)=16:47:14.5, Dec(2000)=$-$01:56:52 }
\begin{tabular}{rcccc}
\hline
Grating &Date       &UT$_{beginning}$ &exptime &airmass \\
\hline
HR11   & 2004-07-20  & 03:50:04  & 2700 &1.41\\ 
HR13   & 2004-04-13  & 07:25:10  & 2700 &1.09\\ 
\hline
\end{tabular}

\label{t:log}
\end{table}

As done with the previous GCs, our targets were selected among stars near
the RGB ridge line and isolated\footnote{All stars were chosen to be free from
any companion closer than 2 arcsec and brighter than $V+2$ mag, where $V$ is
the target magnitude.}. We used the $B,V$ photometry from Sabbi et al. (2006), 
calibrated to the standard Johnson system, for our
target selection. 

The photometric data consist of two dataset. The central region of
NGC~6218 was observed in service mode during 1999 June with 
UT1 of the VLT using FORS1 in the
high resolution (HR) mode. In this configuration the plate scale
is $0\farcs 1/{\rm pixel}$ and the field of view is 
$3\farcm 4 \times 3\farcm 4$. The data consist of exposures of 3, 5 and 7 
seconds in the $B$, $V$, 
$R$ bands, respectively. A complementary set of wide-field $B$ and
$V$ images acquired with the 2.2 m ESO--MPI telescope at ESO (La Silla),
using the Wide Field Imager (WFI) cameras, with a field of view of 
$33'\times 34'$, was retrieved from the ESO data archive. 

The reduction of the HR exposures was carried out using ROMAFOT
(Buonanno etal. 1983, Buonanno \& Iannicola 1986), a package specifically
developed to perform accurate photometry in crowded fields. 
The WFI data were reduced using DAOPHOT II (Stetson 1994). 
The data sets were then matched together, and a final catalog listing the
instrumental $b$, and $v$ magnitudes, and coordinates for all the stars in each
chip was compiled. The Guide Star Catalog (GSCII) was used to search
for astrometric standards in the entire WFI image field of view. Several
hundreds of astrometric GSCII reference stars were found in each chip,
allowing us an accurate absolute position of the detected stars. By using a
sample of bright stars (lying either in the HR and WFI field of view) as
``secondary astrometric standards'', also the astrometric solution of the
innermost region of NGC~6218 was found. The final instrumental
magnitudes of the HR exposures were transformed to the standard Johnson
photometric system by using the same procedure and the same ten photometric
stars adopted in Paltrinieri et al. (2001). Stars in common between the HR
and WFI images were used to calibrate the WFI catalog on the HR one. A final
catalog completely homogeneous in magnitudes and coordinates was
obtained.

\begin{figure}
\centering
\includegraphics[bb=50 175 440 600, clip, scale=0.69]{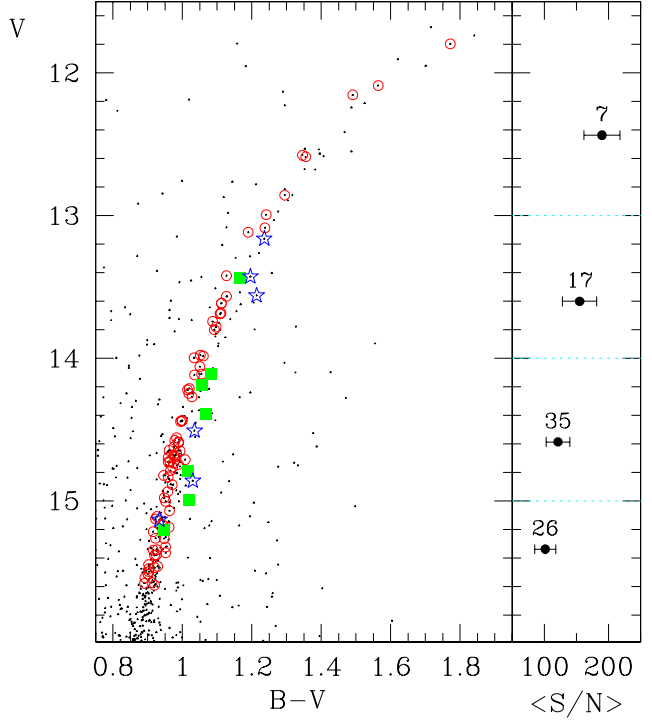}
\caption{$V,B-V$ CMD for NGC 6218 (magnitudes are from Sabbi et al. 2006).
The observed stars are indicated by (red) open
circles if they are bona fide members, (green) filled squares if their RVs
indicate that they are sure non members, and (blue) open star symbols for the
uncertain cases, for which the metallicity has been determined, but have been
excluded from further analysis (see text). The right panel shows the average
S/N ratios of spectra (per pixel) at different magnitude levels; numbers 
indicate how many stars were averaged each time.} 
\label{f:figcmd}
\end{figure}

Table~\ref{t:log} lists information about the two GIRAFFE pointings, while a list of
all observed targets with coordinates, magnitudes and radial velocities (RVs)
is given in Table~\ref{t:coo} (the full table is available
only in electronic form). The $V,B-V$  colour magnitude
diagram (CMD) of our sample is shown in Figure~\ref{f:figcmd}; our targets
range from about $V$= 11.8 to 15.6, i.e. from near the RGB 
tip to more than 1 magnitude below the RGB bump. 
A few field stars (on the basis of their radial velocities) are indicated by
different symbols. 
Contamination from stars on the asymptotic giant branch (AGB) may be
possible for colors redder than $B-V \sim 1.2$; however, this is not of concern
for our analysis, because a priori AGB stars are about 10\% of the RGB stars
and a posteriori the very small scatter in derived abundances supports the
reliability of the adopted atmospheric parameters, including the adopted
stellar mass (appropriate for first ascent red giant branch stars).

We used the 1-d, wavelength calibrated spectra as reduced by the dedicated
Giraffe pipeline (BLDRS  v0.5.3, written at the Geneva Observatory, see
{\em http://girbldrs.sourceforge.net}). Radial velocities (RVs) were 
measured using the {\sc IRAF}\footnote{
IRAF is distributed by the National Optical Astronomical
Observatory, which are operated by the Association of Universities for
Research in Astronomy, under contract with the National Science
Foundation } package {\sc FXCORR} on appropriate templates and are
shown in Table~\ref{t:coo}.
We have noticed a systematic difference between the RVs of stars
observed with both gratings (on average ${\rm RV(HR11)-RV(HR13)}\sim-2.6$ km~s$^{-1}$).
This is completely irrelevant from the point of view of the abundance
analysis and is most probably due to small zero point errors in the
wavelength calibration. We have measured an average shift of +0.02 \AA \ 
(i.e., about $+1$ km~s$^{-1}$) for the two auroral emission lines of [O {\sc i}] at 
6300 and 6363 \AA, but no such check was possible for grating HR11.
Correcting the observed RVs of HR13 for the above mentioned shift and
eliminating the sure and possible non members, the average cluster 
heliocentric  velocity  is $-42.5$ (rms=3.2) \,km s$^{-1}$, in very good
agreement with the value ($-42.2\pm0.5$) tabulated in the updated catalog of
Harris (1996). 

Seven stars were excluded from this average and from further analysis 
as non-members based on the RV: these objects differ by more than 15$\sigma$
from the cluster average.

Furthermore, for one object (star 100010) we found very different RVs in the 
two pointings; this is most likely due to a misidentification of the target in one of
the two pointings. Given its dubious status, we dropped this star from our
sample too. Other objects were discarded from the sample after the abundances
analysis because their metallicity resulted much higher than the
cluster average. Most probably, these are disk objects with RVs similar to the 
cluster's value. To check this possibility we used the Besan\c con
galactic model (Robin et al. 2003) and found that a relevant fraction
of field stars in the direction of NGC~6218 has a RV completely compatible
with the one of the cluster.

Not all the stars were observed in both gratings, since we also aimed to target
up to 14 stars per cluster with the 7 UVES fibers; hence the GIRAFFE fiber
positioning was slightly different between the two pointings.
On a grand total of 79 different stars observed and {\it bona fide} cluster
members, we have 53 objects with spectra for both gratings, 11 with only HR11
observations and 15 with only HR13 observations. We were able however to measure
Na abundances for all target stars but one, because the Na doublet at
6154-6160~\AA\ falls into the spectral range covered by HR13, whereas we could
expect to measure O (or to determine a limit) only for a maximum number of 68
stars.

\begin{table*}
\centering
\caption{List and relevant information for the target stars observed in
NGC~6218. ID, $B$, $V$ and coordinates (J2000) are taken from Sabbi et al.
(2006); $J$, $K$ are from the 2MASS catalog; the RVs
(in km s$^{-1}$) from both gratings are heliocentric. The complete Table is
available electronically; we show here a few lines for guidance. Star 1 to 85
are members (or doubtful members, indicated by NM~ in Notes), star 86 to 92 are
non members based on the RV. A few stars excluded from the analysis are
indicated by an '*' in Notes.}
\begin{tabular}{rrccccccrrl}
\hline
\hline
Nr   &ID      &RA           &Dec         &$B$    &$V$    &$J$    &$K$	 &RV(HR11)  &RV(HR13)&Notes        \\
\hline
   1 &100001  &251.9351289  &-1.8583407  &15.544 &14.509 &12.518 & 11.925 &   -34.55 & -31.84 &  HR11,HR13,*  \\
   2 &100005  &251.9152836  &-1.9122903  &15.890 &14.860 &12.791 & 12.151 &   -35.94 & -33.03 &  HR11,HR13,*  \\
   3 &100010  &251.9115453  &-1.8639458  &16.066 &15.132 &13.215 & 12.673 &   -47.28 &  36.65 &  HR11,HR13,*  \\
   4 &100017  &251.9178348  &-1.9572735  &16.114 &15.170 &13.184 & 12.511 &   -40.52 &        &  HR11	    \\
   5 &100021  &251.9170197  &-1.9697241  &16.261 &15.345 &13.386 & 12.699 &	     & -47.96 &  HR13	    \\
   6 &100027  &251.9258546  &-1.8851002  &16.511 &15.591 &13.610 & 12.963 &   -41.85 & -39.33 &  HR11,HR13  \\
   7 &101677  &251.9417972  &-1.9880637  &14.399 &13.163 &10.813 & 9.933  &   -24.28 & -21.34 &  HR11,HR13, NM?,*  \\
   8 &101682  &251.8963944  &-1.8638070  &14.624 &13.428 &11.085 & 10.295 &	     & -57.05 &  HR13, NM?,*   \\
   9 &200004  &251.8400920  &-1.9497273  &13.944 &12.589 & 9.958 & 9.068  &   -46.41 &        &  HR11	    \\
  10 &200006  &251.8643031  &-2.0068012  &13.923 &12.577 & 9.977 & 9.096  &	     & -36.58 &  HR13	    \\
\hline
\end{tabular}
\label{t:coo}
\end{table*}

\section{Atmospheric parameters and analysis}

\subsection{Atmospheric parameters}

Temperatures and gravities were derived as described in Paper I and Paper II; 
along with the derived atmospheric parameters and iron
abundances, they are shown in Table \ref{t:atmpar} (completely available only
in electronic form).
We obtained T$_{\rm eff}$'s and bolometric corrections B.C. for our stars from
$V-K$ colours, where $K$ was taken from the Point Source
Catalogue of 2MASS (Skrutskie et al. 2006) and transformed to
the TCS photometric system, as used in Alonso et al. (1999).
We employed the relations by Alonso et al.
(1999, with the erratum of 2001). We adopted for NGC~6218 a
distance modulus of $(m-M)_V$=14.02, a reddening of $E(B-V)$ = 0.19,  an input
metallicity  of [Fe/H]$=-1.37$  
(Harris 1996), and the relations  $E(V-K) = 2.75 E(B-V)$, $A_V =
3.1 E(B-V)$, and $A_K = 0.353 E(B-V)$ (Cardelli et al. 1989). 

As done for NGC~6752 (Paper II), the final adopted $T_{\rm eff}$'s were
derived from a relation between  $T_{\rm eff}$ (from $V-K$ and the Alonso et
al. calibration) and $V$ based on 79 "well
behaved" stars (i.e., with magnitudes in all the four filters and lying on the
RGB). This procedure was adopted in order to decrease the scatter in abundances
due to uncertainties in temperatures, since magnitudes are much more reliably
measured than colours.
The assumptions behind this approach are discussed in Paper II and they will be
not repeated here.

Surface gravities log $g$'s were obtained from the apparent magnitudes, the 
above discussed effective temperatures and distance modulus, and the 
bolometric corrections from Alonso et al. (1999), assuming  masses of 
0.85 M$_\odot$ and $M_{\rm bol,\odot} = 4.75$ as the bolometric magnitude 
for the Sun.

We eliminated trends in the relation between abundances of Fe {\sc i} and
expected line strength (see Magain 1984) to obtain values of the
microturbulence velocities $v_t$s.

\subsection{Equivalent widths and iron abundances}

Line lists, atomic parameters and reference solar abundances are from Gratton
et al. (2003). Equivalent widths ($EW$s) were measured as described in detail
in  Bragaglia et al. (2001) with the same procedure adopted in Paper I and
Paper II for the  definition of the local continuum around each line. This is a
crucial step at the limited resolution of our spectra, especially for the
coolest targets. 
According to the strategy illustrated in Paper II, we checked the
reliability of $EW$s measured on the GIRAFFE spectra by performing a comparison
with those measured on the high-resolution UVES/FLAMES spectra of 9 stars
observed in both configurations. For NGC 6218 $EW$s from GIRAFFE are  on
average larger than those measured from UVES spectra  by $+0.6\pm0.5$ m\AA\
(rms=10.0 m\AA\ from 371 lines), i.e. in excellent agreement.
Tables of measured $EW$s are only available at the CDS
database.

Final metallicities are obtained by choosing in the Kurucz (1993) grid of model
atmospheres (with the option for overshooting on) the model with the proper
atmospheric parameters whose abundance matches that derived from Fe {\sc i} 
lines.

\begin{figure}
\centering
\includegraphics[bb=120 215 520 700, clip, scale=0.65]{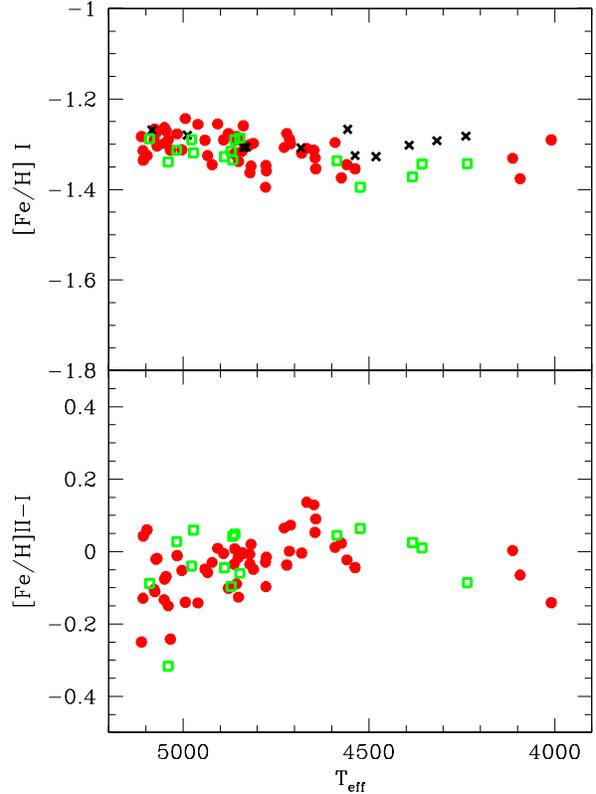}
\caption{Run of [Fe/H] ratio and of the iron ionization equilibrium as a
function of temperatures for program stars in NGC 6218. Symbols and color
coding refer to the setup used: (red) filled circles indicate stars with 
both HR11 and HR13 observations,
(black) crosses for HR11 only, and (green) empty squares for HR13 only.}
\label{f:feteff}
\end{figure}

Average abundances of iron for NGC~6218 are [Fe/H]{\sc i}=$-1.31$ (rms=0.03 
dex, 79 stars) and [Fe/H]{\sc ii}=$-1.35$ (rms=0.08 dex, 68
objects). The very good agreement indicates a self-consistent analysis, since
the ionization equilibrium for Fe is very sensitive to any possible problem in
the abundance analysis. The difference we found is scarcely significant, and
the star-to star scatter in the residuals is very close to the value we might
expect on the basis of errors in the measurements of $EW$s, by considering that 
on average only three lines of Fe {\sc ii} were typically used  in the
analysis.
The distribution of the resulting [Fe/H] values and of the difference between
ionized and neutral iron are shown in Figure~\ref{f:feteff} as a function of
temperature, with stars coded according to the grating(s) they were observed
with. There is no significant trend as a function of temperature over 
a range of almost four magnitudes.

The (small) scatter of the metallicity distribution is discussed 
in Sect.5.

\begin{table*}
\centering
\caption[]{Adopted atmospheric parameters and derived iron abundances in 
stars of NGC 6218; nr indicates the number of lines 
used in the analysis. 
The complete Table is available only in electronic form.
}
\begin{tabular}{rccccrcccrccc}
\hline
Star   &  T$_{\rm eff}$ & $\log$ $g$ & [A/H]  &$v_t$	     & nr & [Fe/H]{\sc i} & $rms$ & nr & [Fe/H{\sc ii} & $rms$ \\
       &     (K)	&  (dex)     & (dex)  &(km s$^{-1}$) &    & (dex)	  &	  &    & (dex)         &       \\
\hline
  100017  &4989 &2.54 &-1.28 &0.81 &  5 &-1.28 &0.076  &     &	  &	  \\	
  100021  &5040 &2.61 &-1.34 &1.10 & 11 &-1.34 &0.168  &  2  &-1.66 &0.163  \\	
  100027  &5112 &2.72 &-1.28 &1.45 & 13 &-1.28 &0.107  &  1  &-1.53 &	  \\	
  200004  &4239 &1.07 &-1.28 &1.66 & 16 &-1.28 &0.149  &     &	  &	  \\	
  200006  &4235 &1.08 &-1.34 &1.28 & 19 &-1.34 &0.126  &  3  &-1.43 &0.114  \\	
  200014  &4317 &1.22 &-1.29 &1.36 & 15 &-1.29 &0.062  &     &	  &	  \\	
  200019  &4357 &1.34 &-1.34 &1.77 & 21 &-1.34 &0.150  &  3  &-1.33 &0.141  \\	
  200024  &4383 &1.39 &-1.37 &1.46 & 20 &-1.37 &0.111  &  3  &-1.35 &0.158  \\	
  200025  &4392 &1.44 &-1.30 &1.44 & 12 &-1.30 &0.095  &     &	  &	  \\	
  200035  &4523 &1.70 &-1.40 &1.36 & 19 &-1.40 &0.093  &  3  &-1.33 &0.127  \\	
 \hline
\end{tabular}
\label{t:atmpar}
\end{table*}

\begin{table*}
\centering
\caption[]{Sensitivities of abundance ratios to variations in the atmospheric
parameters and to errors in the equivalent widths, as computed for a typical 
program star with T$_{\rm eff} \sim 4600$ K. 
The total error is computed as the quadratic sum
of the three dominant sources of error,
T$_{\rm eff}$, $v_t$ and errors in the $EW$s, scaled to the actual errors as
described in the text (Col. 8: tot.1) or as the sum
of all contributions (Col. 9: tot.2)}
\begin{tabular}{lrrrrrrrr}
\hline
\\
Ratio    & $\Delta T_{eff}$ & $\Delta$ $\log g$ & $\Delta$ [A/H] & $\Delta v_t$
&$<N_{lines}>$& $\Delta$ EW & tot.1 & tot.2\\
         & (+50 K)    & (+0.2 dex)      & (+0.10 dex)      & (+0.10 km~s$^{-1}$) & & & (dex)& (dex)  \\
 (1)                & (2)     & (3)      & (4)     & (5)      & (6)& (7)    & (8)  & (9) \\
\\    
\hline
$[$Fe/H$]${\sc  i}  &  +0.055 & $-$0.005 &$-$0.006 & $-$0.019 & 24 &+0.025  &0.039 &0.039  \\
$[$Fe/H$]${\sc ii}  &$-$0.023 &   +0.086 &  +0.020 & $-$0.008 &  3 &+0.071  &0.072 &0.073  \\
\hline
$[$O/Fe$]$          &$-$0.041 &   +0.086 &  +0.036 &   +0.020 &  2 &+0.087  &0.092 &0.093  \\
$[$Na/Fe$]$         &$-$0.018 & $-$0.029 &$-$0.004 &   +0.011 &  3 &+0.071  &0.073 &0.073  \\
\hline
\end{tabular}
\label{t:sensitivity}
\end{table*}

\section{Errors in the atmospheric parameters and cosmic scatter in Iron}

Following the tested procedure of Paper I and II, we estimated 
individual (i.e. star-to-star) errors in the derived abundances.
These errors are those relevant when discussing the internal
spread of abundance within a cluster, the main aim of our study in NGC 6218.

We estimated this kind of uncertainty
by considering the three main error sources, i.e.
errors in temperatures, in microturbulent velocities and in the
measurements of $EW$s. 
We will see that the effects of errors in
surface gravities and in the adopted model metallicity are negligible on the
total error budget.
Hence, in the following we will concentrate on the major error sources.

The first step is to evaluate the sensitivity of the derived abundances 
for Fe, Na and O on the adopted atmospheric parameters.
This was obtained by re-iterating the analysis while varying each time only one of the
parameters; the amount of the variations and the resulting sensitivities are
shown in Table~\ref{t:sensitivity}, Columns from 2 to 5.

This exercise was done for all stars in the sample. The average
value of the slope corresponding to the average temperature ($\sim 4600$ K) in
the sample was used as representative to estimate the internal errors 
in abundances, when combined with realistic estimates of internal (i.e.
star-to-star) errors in the atmospheric parameters, presented in the following.

\paragraph{Errors in temperatures.} 
Like we did for NGC~6752,
we adopted a
final T$_{\rm eff}$ value from a calibration of temperature as a function of
the $V$ magnitude. We adopted this procedure thanks to the very small errors
produced in T$_{\rm eff}$.
The nominal internal error in T$_{\rm eff}$ is estimated to be about 6~K, from the
adopted error in $V$ (a conservatively large value of 0.02 mag, likely 
overestimated, in this magnitude range) and the slope of the relation between 
temperature and magnitude (291 K/mag, along the RGB of NGC 6218).
We would like to emphasize here that it is this internal contribution to affect
the overall error budget when comparing stars in an individual cluster.
Real internal errors in temperatures are likely larger than this nominal 
value, due to a variety of reasons (see discussion in Paper II). However, we
will still adopt this nominal value in the following discussion, because use of
larger values would result in expected rms scatter of abundances much larger
than observed.

\paragraph{Errors in microturbulent velocities.}
We used star 302157 and repeated the analysis changing $v_t$; the error on
the micro-turbulence velocities is estimated by the change on $v_t$ required
to vary the slope of the expected line strength vs abundances relation by
1$\sigma$ value\footnote{This value was derived as the quadratic mean of the 1
$\sigma$ errors in the slope of the relation between abundance and expected
line strength for all stars with more than 15 lines measured.} 
from the original value.
The corresponding internal error is 0.15 km~s$^{-1}$. 

\paragraph{Errors in measurement of equivalent widths.}
In order to estimate this contribution, we selected a subset of 65 stars with 
more than 15 measured Fe lines. The average rms scatter (0.123 dex) in Fe
abundance for these stars, divided by the square root of the typical average 
number of measured lines (24), provides a typical internal error of 0.025 dex. 

\paragraph{}
Once we have derived the internal errors we may compute the final errors in
abundances; they depend on the slopes of the relations between the variation
in each given parameter and the abundance. 
When combined with sensitivities of Table~\ref{t:sensitivity}, the derived
individual star errors for Fe amount to 0.007 dex and 0.029 dex due
to the quoted uncertainties in T$_{\rm eff}$ and $v_t$.

Summing in quadrature these contributions with the impact of errors in $EW$s 
we may evaluate the expected scatter in [Fe/H] due to the most relevant 
uncertainties in the analysis. We 
derive $\sigma_{\rm FeI}$(exp.)=$0.039 \pm 0.005$ dex (statistical error). The
inclusion of contributions due to uncertainties in surface gravity or model
metallicity does not alter our conclusions. 

Total errors, computed using only the dominant terms or including all the
contributions, are reported in  Table~\ref{t:sensitivity}, in Cols. 8 and 9
respectively, for iron and for the other two elements measured in this paper.

On the other hand, the observed scatter is formally (slightly) lower:
$\sigma_{\rm FeI}$(obs.)=$0.033 \pm 0.004$ (statistical error), obtained as
the rms scatter of abundances from Fe {\sc i} lines.
Within the statistical uncertainties this difference is not
significant, and might indicate that the errors are slightly overestimated.
The conclusion from our large dataset is that 
the observed star-to-star rms scatter in Fe abundances in NGC~6218 is no
more than 8\%.

As for NGC~6752 (see Paper II), we do not find any measurable intrinsic spread
in metallicity in NGC~6218, so we conclude that this cluster is  very
homogeneous as far as the global metallicity is concerned. The stars in our
sample are well distributed over almost 5 half-mass radii: the inference is
that the heavy metal content of the pristine material from which the stars
formed was very well mixed at the time of stellar formation in NGC 6218. As
already noted in Paper II, this is a strong constraint for
any model of cluster formation.

Finally, systematic errors are not relevant within a single cluster, but could
be used when comparing different GCs in the present study.
Briefly, the adopted relation between temperatures and magnitude $V$  is based
on T$_{\rm eff}$ from dereddened $V-K$, through the calibrations by Alonso et
al. Hence, errors in the adopted reddening translate into systematic errors in
derived temperatures. 

The residual slope of the relation of Fe I abundances with excitation potential
is $-0.004\pm 0.003$ dex/eV (rms=0.025 for individual stars). 
Following the approach discussed in Paper II (sect. 5.1), this difference
implies an error of 0.005 mag in $E(B-V)$ or 0.014 mag in $E(V-K)$. When
combined with the slope $\sim 1009$ K/mag of the T$_{\rm eff}$ vs $V-K$
relation, we obtain a systematic uncertainty of 14 K for the cluster,
increasing up to 25 K if we add in quadrature another contribution of a 0.02
mag error in the zero point of $V-K$ colour.

Errors in surface gravity might be obtained by propagating uncertainties in
distance modulus (about 0.1 mag), stellar mass (a conservative 10\%) and the
above error of 25 K in effective temperature. The quadratic sum results into a
0.057 dex of error in $\log g$. 
The systematic error relative to the microturbulent velocity $v_t$, a quantity
derived from our own analysis, was divided by the square root of the number of 
the observed stars, and is estimated in 0.017 km/s.

By using again the sensitivities evaluated in Table~\ref{t:sensitivity}, we can
translate the contributions of systematic errors in T$_{\rm eff}$, $\log g$ and
$v_t$ to the metallicity. Summing in quadrature these contribution with the
statistical error of individual abundance determinations, 0.004 dex, we end up
with  a total systematic error of 0.028 dex.

Note that these errorbars do not include scale errors, due e.g. to errors in
the temperature scale by Alonso et al., to the adoption of the Kurucz model
atmosphere grid, to departures from LTE in line formation, et cetera. However,
these scale errors should have similar effects for stars in different clusters,
insofar the atmospheric parameters are also similar.
Moreover, second-order effects, e.g. in temperature, cannot be very large,
since this would give a much larger spread than the one we observe.

Hence, on the scale we are defining throughout this series of papers, the metal
abundance of NGC 6218 is [Fe/H]$=-1.31\pm 0.004\pm 0.028$ dex, where the first
error bar refers to the individual star errors and the second one is relative
to the cluster or systematic error.

\section{Results: the Na-O anticorrelation}

We derived abundances of O and Na from measured $EW$s. 
Abundances of Na could be derived for all stars but one; depending on the
setup  used, at least one of the Na~{\sc i} doublets at 5672-88~\AA\ and at
6154-60~\AA\ is always  available. Derived average Na abundances were
corrected for effects of departures from the LTE assumption according to the
prescriptions by Gratton et al. (1999).

Oxygen abundances are obtained from the forbidden [O {\sc i}] lines at 6300.3
and 6363.8~\AA; the former has been cleaned from telluric contamination by
H$_2$O and O$_2$ lines, following the procedure of Paper I and II.
Neither
CO formation nor the high excitation Ni {\sc i} line at 6300.34~\AA\
are a source of concern: low C abundances are 
expected, given also the rather high temperatures of our stars, and
the contribution of Ni to the measured $EW$ is negligible (see also Paper II).

In Table~\ref{t:abunao} we list the abundances of O and Na (the complete table
is available only in electronic form). For O we distinguish between actual
detections and upper limits; the number of measured lines and the rms values
are also indicated. We did not try to remove cosmic rays, since the removal can
be difficult and not very precise with a single exposure in each wavelength 
region. However, all the O measures were interactively checked on the HR13 
spectra. Moreover, the Na measurements were interactively checked in all
cases where discrepancies between abundances from the 2 to 4 different lines 
were present.  

\begin{figure}
\centering
\includegraphics[bb=20 140 550 700, clip, scale=0.47]{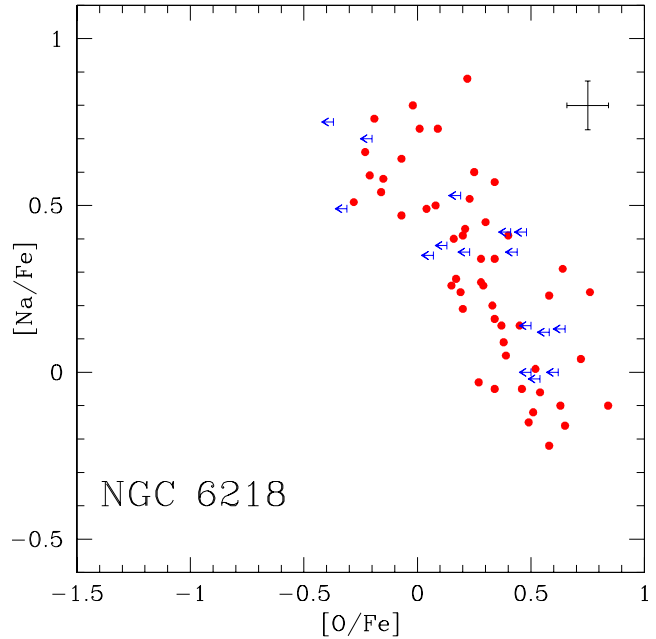}
\caption{[Na/Fe] ratio as a function of [O/Fe] for red giant 
stars in NGC~6218. Upper limits in [O/Fe] are indicated
as blue arrows. The (random) error bars take into account the uncertainties in
atmospheric parameters and $EW$s.}
\label{f:anti}
\end{figure}

The [Na/Fe] ratio as a function of [O/Fe] ratio is displayed  in
Figure~\ref{f:anti} for each of the red giant stars with  both O and Na
detections (filled dots) and with upper limits for [O {\sc i}] (arrows). 
The overall random errors in O and Na, as due
to the contribution of errors in the adopted atmospheric parameters and
measurements of $EW$s, are also indicated.

No comparison with literature data is possible, since this is the first work to
present Na and O abundances for RGB stars in this cluster.

The Na-O anticorrelation in NGC~6218 appears to be normal, without features
worthy of note. In Figure~\ref{f:anti} there is a good number of stars that
appear to have the typical composition of field halo objects, high [O/Fe] and 
slightly less than solar [Na/Fe] ratios. In the O-poor regime, the
anticorrelation is 
not very much extended toward extreme values, although the presence of upper
limits in oxygen in the Na-rich regime hints for the possible existence of a
few O-poor stars in this cluster. The overall appearance of the anomalies in
Na, O is similar to that observed in other metal-intermediate GCs, such as, for
instance, M~3 (Sneden et al. 2004). 

On the other hand, we note that NGC~6218 deserves the full status of
second-parameter cluster.

In fact, let us consider the issue of the HB morphology. M~3 is the classical
template of a globular cluster,
with stars uniformly distributed on a red HB, within the
instability strip, and in a blue HB (HB parameter HBR = (B-R)/(B+V+R) = 0.08). 
NGC~6218 has almost the same  metallicity ([Fe/H]$=-1.31$ dex compared to
$-1.34$ dex for M~3, on the homogeneous scale by Carretta and Gratton 1997) but
shows a very different HB morphology (HBR = 0.92): there are no red or variable HB
stars and only the blue part of the HB is populated, although 
the blue HB is not extended as those of other intermediate-metallicity clusters
such as  NGC~6752 or M~13 (HBR = 1.0 and 0.97, respectively).

Regarding the chemical anomalies, the Na-O anticorrelation in NGC~6218 seems 
rather similar to the one of M~3, but is very different from that of M~13, while
the similarity of HB populations is the other way around. It is a direct
inference that GCs showing the same or very similar {\em first parameter}
(i.e., the metallicity) driving the HB morphology show not only the classical
{\em second-parameter} effect in the distribution of stars along the HB, but
also significant differences in how the chemical anomalies in their light
elements show up along the giant branch. 

It should obviously be noticed that NGC6218 might be
substantially older than M3, possibly explaining the different mean colour
of its HB. As a matter of fact, de Angeli et al. (2005) indeed found a
rather large age difference between these two clusters, M3 being much
younger than NGC6218. Using e.g. the data from ground based observations
(available for both clusters) and the Carretta and Gratton (1997)
metallicity scale, the relative ages found by de Angeli et al. are
$0.76\pm 0.04$\ for M3 and $0.95\pm 0.09$\ for NGC6218. A similar age
difference of some 2-3 Gyr should be large enough to explain the different
mean colour of the HB.

\begin{table*}
\centering
\caption[]{Abundances of O and Na in NGC~6218. [Na/Fe] values are corrected for
departures from LTE. HR is a flag for the grating used 
(1=HR13 only, 2=HR11 and HR13, 3=HR11 only) and lim is a flag discriminating
between real detections and
upper limits in the O measurements (0=upper limit, 1=detection). The complete
Table is available only in electronic form.}
\begin{tabular}{rcccccccc}
\hline
 Star     & nr &[O/Fe]& rms  &  nr &[Na/Fe] &rms &HR & lim \\    
\hline                                                           
   100017 &    &       &      &  2 & 0.19  &0.00 & 3 &  \\ 
   100021 &  1 &  0.48 &      &    &	   &	 & 1 & 0\\ 
   100027 &    &       &      &    &	   &	 &   &  \\ 
   200004 &    &       &      &  2 &  0.77 &0.10 & 3 &  \\ 
   200006 &  2 &  0.45 &0.05  &  2 &  0.14 &0.07 & 1 & 1\\ 
   200014 &    &       &      &  2 &  0.24 &0.02 & 3 &  \\ 
   200019 &  2 &  0.34 &0.02  &  2 &  0.57 &0.04 & 1 & 1\\ 
   200024 &  2 &  0.33 &0.08  &  2 &  0.20 &0.00 & 1 & 1\\ 
   200025 &    &       &      &  2 &  0.01 &0.06 & 3 &  \\ 
   200035 &  2 &  0.34 &0.03  &  2 &  0.16 &0.19 & 1 & 1\\ 
\hline
\end{tabular}
\label{t:abunao}
\end{table*}

\begin{figure}
\centering
\includegraphics[bb=55 185 400 700, clip, scale=0.67]{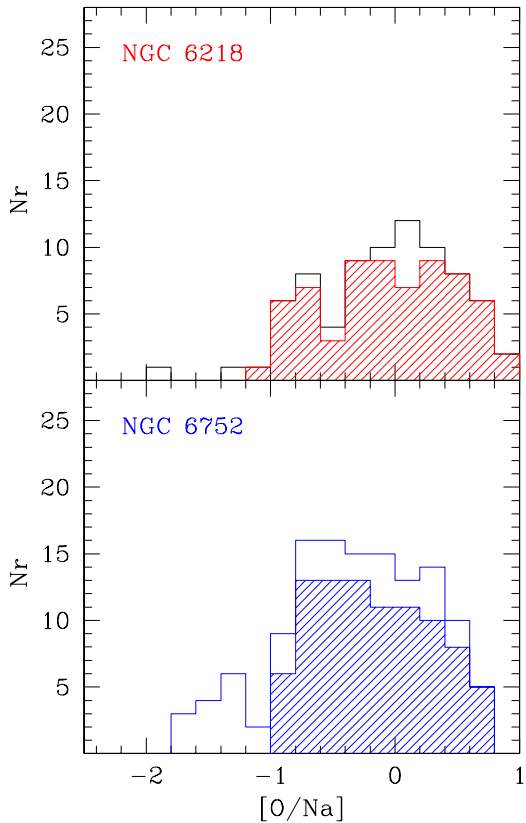}
\caption{Upper panel: distribution function of the [O/Na] ratios along 
the Na-O anticorrelation in NGC~6218. The dashed area is the frequency
histogram referred to actual detection or limits of O in stars, whereas the
empty histogram is obtained by using the global anticorrelation relationship
derived in Paper I to obtain O abundances also for stars with no observations
with HR13. Lower panel: the same for NGC 6752 (Paper~II).}
\label{f:histom62m67}
\end{figure}

\begin{figure}
\centering
\includegraphics[scale=0.67]{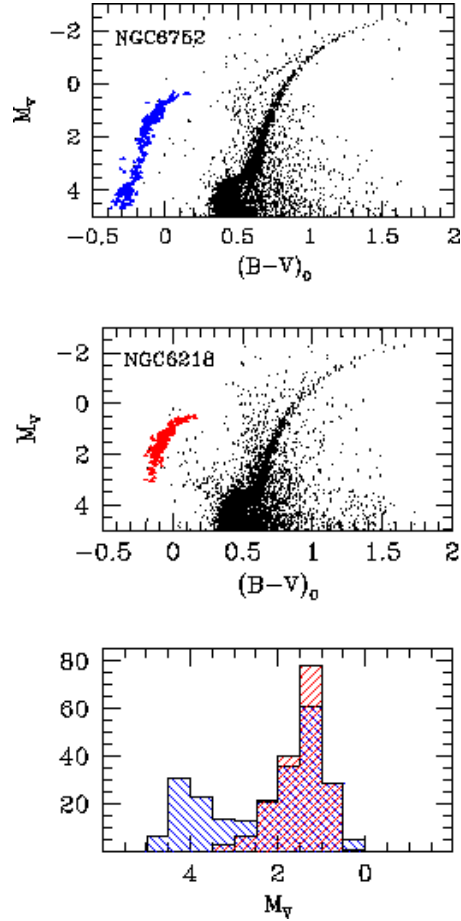}
\caption{CMDs of NGC~6752 (Momany et al. 2004, upper panel)
and of NGC~6218 (Sabbi et al. 2006, middle panel), with HB stars shown as
heavier symbols. In the lower panel, the
luminosity distributions of stars along the horizontal branches of the two
clusters (in red for NGC~6218, in blue for NGC~6752) are shown.}
\label{f:histom62hb2}
\end{figure}

\begin{figure}
\centering
\includegraphics[bb=36 161 310 700, clip, scale=0.67]{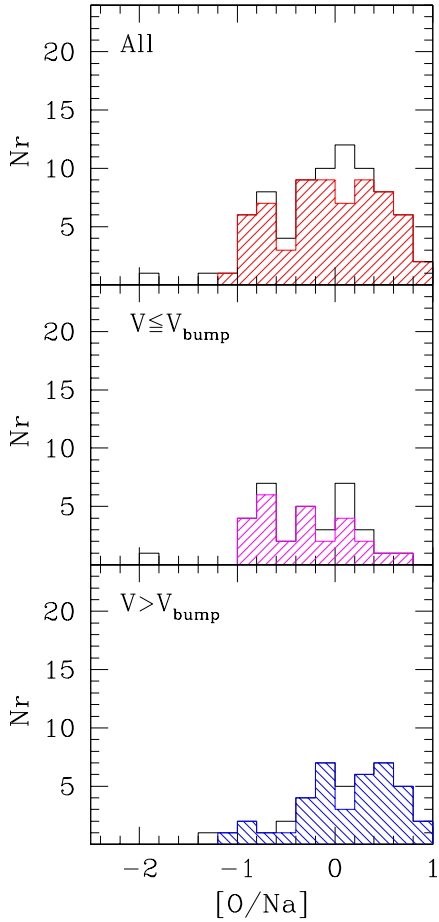}
\caption{Distribution function of the [O/Na] ratios along 
the Na-O anticorrelation in NGC~6218. The upper panel shows the total
distribution, middle and lower panels are restricted to stars brighter and
fainter than the magnitude of the RGB bump ($V=14.60$), respectively.
Symbols are as in the previous figure.}
\label{f:histo3m62}
\end{figure}

\begin{figure}
\centering
\includegraphics[bb=40 156 566 689, clip, scale=0.47]{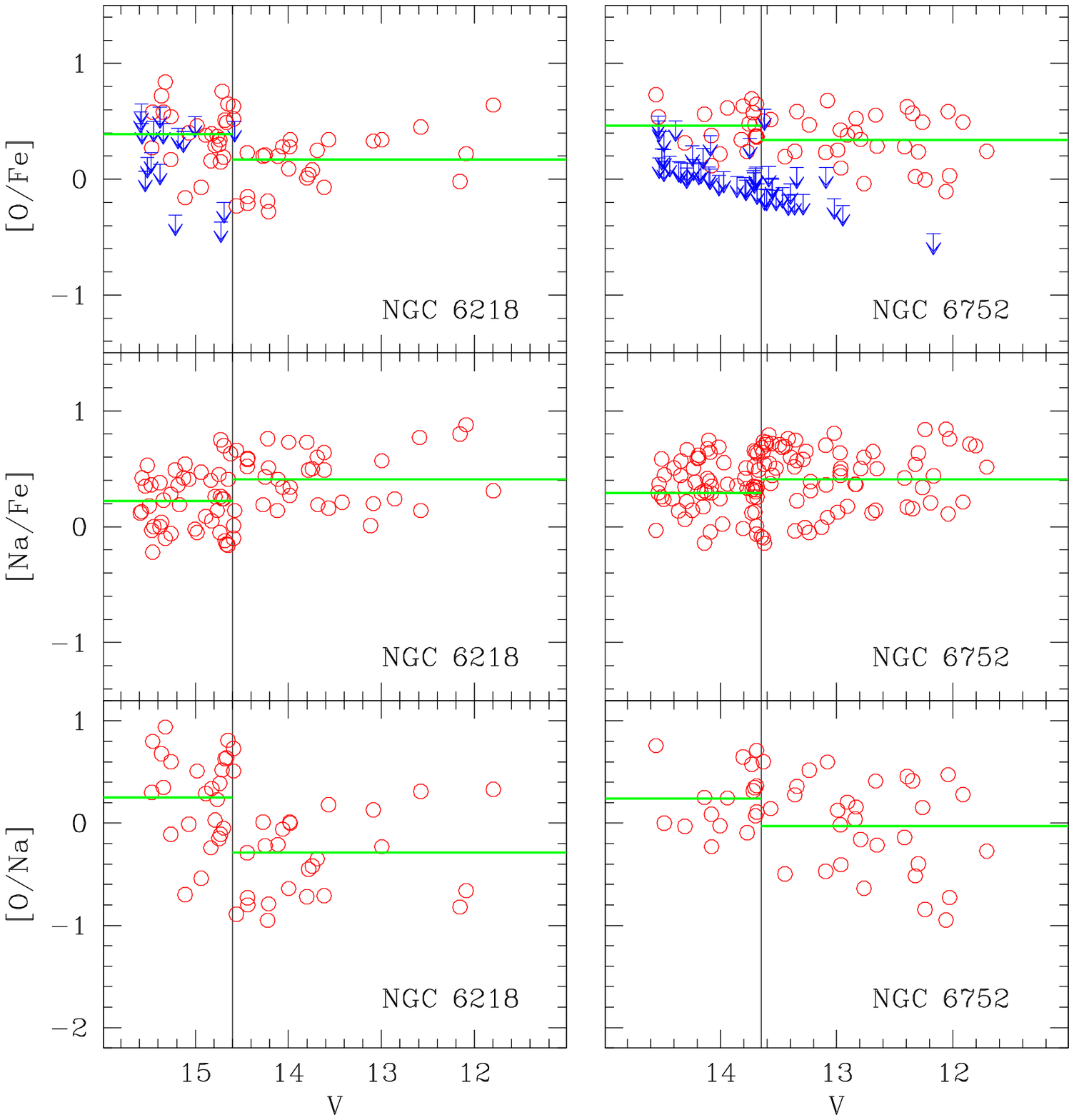}
\caption{Run of the [O/Fe], [Na/Fe] and [O/Na] ratios as a function of the $V$
magnitude for stars analyzed in NGC 6218 (left panels, present work) and in NGC
6752 (right panels, Carretta et al. 2006). The vertical lines mark the
magnitude of the RGB-bump in each cluster ($V=14.60$ and 13.65, respectively. 
The horizontal lines indicate the average ratio for stars fainter and brighter
than the bump (see Table~\ref{t:vona}; they are computed for [O/Fe] and [O/Na]
by using only actual detections in O, excluding stars with only upper limits.}
\label{f:vona6267}
\end{figure}

A more direct and homogeneous comparison is possible:
in NGC 6752, another GC with quite similar metallicity ([Fe/H]$=-1.46$ dex, 
Paper II) and age (see de Angeli et al. 2005), we analyzed a large number
of stars with homogeneous procedures, atomic parameters, packages. In both
clusters we sampled a range of almost four magnitudes along the red giant
branch, and in both GCs the progeny of RGB stars ends up on a HB only populated
on the blue side of the instability strip, with no red or variable HB stars.

The distribution function of stars along the Na-O anticorrelation in NGC~6218
is shown in the upper panel of Figure~\ref{f:histom62m67}.
The dashed area shows the distribution obtained by using only actual
detections or carefully checked upper limits; the empty histogram is derived
by following the overall Na-O anticorrelation, as obtained in Paper I,
in order to get [O/Fe] values even for stars with no observations in HR13.
The bottom panel reproduces the same distribution for NGC~6752, taken from
Paper II. 
The two distributions look different, and this visual impression
is confirmed by a Kolmogorov-Smirnov test. In NGC~6218 the number of O-poor
and Na-rich stars (with ratio [O/Na]$\leq 0$) is roughly comparable to the
number of O-rich stars, at variance with the case of NGC~6752, where there is a
preponderance of stars with O-poor/Na-rich composition. 

This difference in the ratios of chemical anomalies is in some way mirrored by
differences in the HB morphology. As shown in Figure~\ref{f:histom62hb2}, in
NGC~6752 the hot blue tail of the HB is populated down to very faint
magnitudes, reaching the level of the turn-off point of the Main Sequence.
On the contrary, in NGC~6218 the hottest tail of the blue HB is completely
absent.

We conclude that these comparisons show that the mechanism generating the
chemical anomalies in GCs is also a viable candidate as one of the possible
several ``second parameters" determining the morphologies of HBs in GCs, 
although other parameters like e.g. the cluster age, should also be taken 
into account.

\section{The RGB-bump in the luminosity functions of He-poor and He-rich stars}

\begin{table*}
\centering
\caption[]{Mean [O/Fe], [Na/Fe] and [O/Na] values for stars brighter and
fainter than the magnitude of the RGB bump for the two clusters NGC~6218 
and NGC~6752. The difference is in the sense faint minus bright.}
\begin{tabular}{lccc}
\hline
           &$V>V_{bump}$  &$V<V_{bump}$   &diff \\
\hline
\multicolumn{4}{c}{NGC~6218}\\
${\rm [O/Fe]}$   & +0.39 $\pm$0.05  &  +0.17 $\pm$0.05 &   +0.22 $\pm$0.07 \\
${\rm [Na/Fe]}$  & +0.22 $\pm$0.04  &  +0.41 $\pm$0.04 & $-$0.19 $\pm$0.06 \\
${\rm [O/Na]}$	 & +0.25 $\pm$0.09  &$-$0.29 $\pm$0.09 &   +0.54 $\pm$0.13 \\
\multicolumn{4}{c}{NGC~6752}\\
${\rm [O/Fe]}$   & +0.46 $\pm$0.04  &  +0.34 $\pm$0.04  &  +0.12 $\pm$0.06 \\
${\rm [Na/Fe]}$  & +0.29 $\pm$0.04  &  +0.41 $\pm$0.04  &$-$0.12 $\pm$0.06 \\
${\rm [O/Na]}$	 & +0.24 $\pm$0.07  &$-$0.03 $\pm$0.08  &  +0.27 $\pm$0.11 \\
\hline
\end{tabular}
\label{t:vona}
\end{table*}

NGC~6218 is the second cluster in the present series in which we observed
stars well below the magnitude of the RGB-bump, $V=14.60$ from our data. 
As done for NGC~6752,
we separate our dataset plotting in the middle and lower panels of
Figure~\ref{f:histo3m62} the distributions along the Na-O anticorrelation
for stars brighter and fainter than the RGB bump, respectively.
At variance with the case of NGC~6752, the two subsamples in NGC~6218 show
distinctively different distributions. Again, this finding is confirmed by
a statistical (Kolmogorv-Smirnov) test.
From  these distributions the abundances appear shifted
towards lower [O/Fe] and higher [Na/Fe] ratios for stars brighter than the
magnitude of the RGB-bump.

Is this effect the manifestation of the so-called evolutionary mechanisms at
work?
The presence of a well defined Na-O anticorrelation before the bump supports
the primordial origin for the chemical anomalies, already amply accepted
(Gratton et al. 2004).
On the other hand, how can we explain why the two 
distributions are  different? There is no good mechanism to invoke in order to
produce the appearance of  such an evolutionary effect. 
M~13 is the only other cluster in which a shift towards more Na-rich
and O-poor compositions ascending along the RGB had been proposed (Pilachowski
et al. 1996). However, recently Sneden et al. (2004) reached the conclusion
that an evolutionary origin for these anomalies is doubtful and must be
rejected also in the case of M~13.

An evolutionary contribution to the variations of lighter elements such as
C and N is expected (and observed, see Smith and Martell 2003). 
On the other hand, it is well known that Pop. II low mass giants never reach
the central temperatures required to activate the proton-capture reactions
required to alter the content of heavier nuclei such as O, Na, Al, Mg (see
Gratton et al. 2001). Hence, no dependence of abundances for these elements
would be  expected as a function of luminosity: apparently this seems at odds
with what we observe in NGC~6218.

To assess further on the reality of this effect we show in
Figure~\ref{f:vona6267} the run of the [O/Fe], [Na/Fe] and [O/Na] ratios as a
function of the $V$ magnitude both for NGC~6218 (left panels) and NGC~6752
(right panels). The vertical lines in the Figure indicate the RGB bump level,
whereas the horizontal lines indicate the average values we found for the two
subsamples in each cluster. These values are listed in Table~\ref{t:vona} and
are computed for [O/Fe] (and [O/Na]) by using only actual measures, 
excluding any stars with only an upper limit in O. 

The Na abundance ratios may be regarded as a more robust indicator than
O ratios, because of the presence of limits in O abundances; hence
we will concentrate on the former ones. 

While the differences in mean abundance ratios in NGC~6752 are not 
statistically 
significant, in NGC~6218 they are and we can notice a rather clear 
discontinuity at the bump level. This is more a
step-like pattern than a continuous increase or decrease in Na and O,
respectively, as would be expected for an evolutionary effect, 
like a supposed extra-mixing
scenario where O-depletions and Na-enhancements become more marked as the
stars climb along the RGB (but only {\em after} the RBG-bump as in the field
analogs).

The overall
pattern could be well described as a mild trend as a function of the $V$
magnitude (or effective temperature) superimposed to a step-like discontinuity
exactly at the luminosity level of the bump in the luminosity function (LF) of
the RGB. This discontinuity is more clear in NGC~6218 than in NGC~6752.
What is the explanation of this apparently unexpected behavior? 

The first inference is thus that the RGB-bump is involved in some way: it is
unlikely that this major alteration is observable exactly at $that$
magnitude just by chance.
The second certainty is that it cannot be an evolutionary effect, because the
Na-O anticorrelation is present well below this luminosity.

The gentle, overall trend (an increase of about 0.1 dex in [Na/Fe] over 
almost 4 magnitudes in $V$) is likely due to the combination of two factors:
\begin{itemize}
\item[a)] small uncertainties in the corrections for departures from the LTE
assumption (Gratton et al. 1999). These corrections are a function of
temperature and we may expect that associated errors show up 
particularly in NGC~6218 because of the rather small internal errors in the
analysis and the large range of T$_{\rm eff}$'s sampled; 
\item[b)] it is well known (e.g. Dalle Ore 1993) that the model atmospheres by
Kurucz (1993) might not provide an accurate reproduction of the correct
temperature stratification T($\tau$) with respect to the real stars in the
cool, low-gravity and metal-poor regime. In particular, the actual temperature
gradient should be steeper, as required to better reproduce the observations.
This would also be consistent with the 3-D model atmospheres including
adiabatic expansion of the convective cells (see Asplund et al. 1999). A
steeper gradient implies in turn that elements with low excitation potential
present lines that appear stronger than expected at low temperatures.
Hence, Na abundances could be slightly overestimated.
\end{itemize}

A fair estimate of the total trend due to the above effects at the metallicity 
of NGC~6218 is about 0.1 dex.

However, we found that apart from this trend there is a quite real effect, that
can possibly be explained by 
{\em different RGB-bumps on the LFs for the He-poor and He-rich components of
the RGB in NGC~6218}. 

Very recently, Salaris et al. (2006) computed stellar models and isochrones
including the metal mixtures observed in real GC stars due to the 
abundance anticorrelations. They investigated the 
effect of different degrees of chemical anomalies on cluster sequences and LFs.
In particular, Salaris et al. studied the changes induced in the LF along 
the RGB by a varying He content of the polluting material, finding that the
largest effect is at the RGB-bump region: at fixed age and metallicity, 
the bump becomes brighter as the He content increases, and the difference may
be about 0.1 mag (of the same order of the intrinsic width of the bump
itself), depending on the relative weights of the components concurring to
shape the bump in the LF.

How could this finding help to explain the observed feature in NGC~6218? The
main issue here is the theoretical prediction that the LFs of He-poor and
He-rich stars are different. Far from the bump region
we cannot notice any relevant effect. However, in the limited region
bracketing $V_{\rm bump}$, stars with high Y  should have
a RGB-bump brighter than the one of stars with normal Y content, due to the
higher ratio Z/X.

Observationally these He-rich and He-poor components translate into 
O-poor/Na-rich and O-rich/Na-poor groups, because of the pattern due to
pollution from matter processed in H-burning at high temperature. Thus, at the
somewhat brighter magnitude of the RGB-bump of He-rich stars we should expect
to find predominantly O-poor stars (hence a higher mean value of Na). On the
contrary, at the fainter magnitude of the bump in the LF of He-poor stars we
should observe predominantly O-rich stars (with a lower mean value of Na).

This is how we may interpret the run of [Na/Fe] in NGC~6218 
(Figure~\ref{f:vona6267}, left panels), but another test can be made.
We adopt Na as a preferred indicator, because it has no upper limits but
only measured abundances available
for all stars in the sample. By plotting the distribution of [Na/Fe] ratios as
a function of magnitude, we should see a "dip" just before the average
RGB-bump level, in
correspondence to the peak in the LF of He-poor stars, followed by the bump at
brighter magnitude due to the peak in the distribution of He-rich/Na-poor
stars.

\begin{figure*}
\centering
\includegraphics[bb=30 430 570 700, clip, scale=.8]{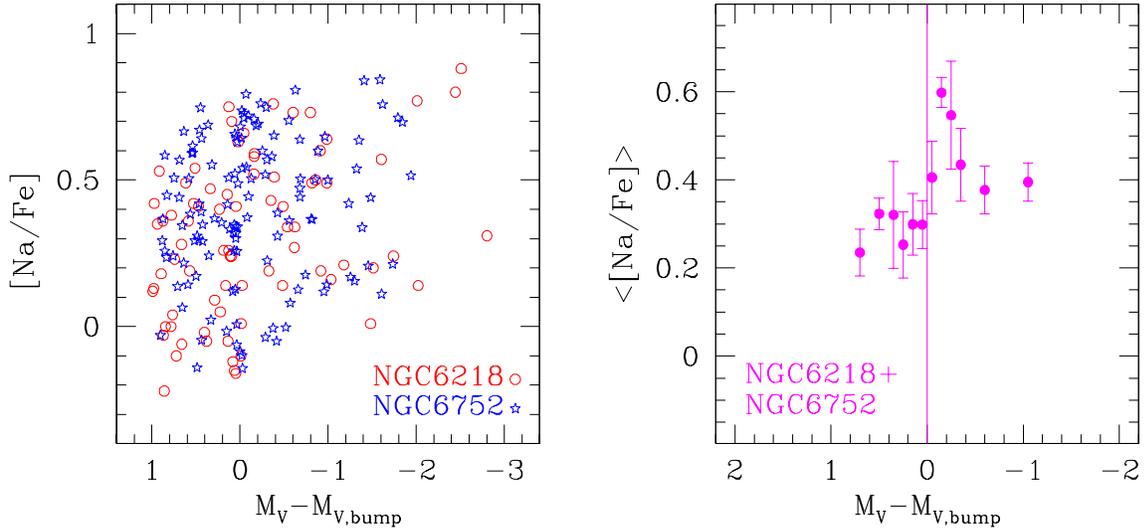}
\caption{Left panel: [Na/Fe] ratios for stars in NGC~6218 (present
work) and NGC~6752 (Paper II) as a function of the absolute magnitude $M_V$,
normalized to the respective RGB-bump absolute magnitudes. Right panel:
distribution of average [Na/Fe] ratios.
Error bars are from the Poisson statistics.}
\label{f:hebump}
\end{figure*}

We try to see if this is true in Figure~\ref{f:hebump}. To
alleviate the problem of small number statistics, we summed the samples 
in NGC~6218 (the present work) and in NGC~6752 (Paper II). We feel
justified in this  
approach because the two clusters were studied by using a very homogeneous
procedure.
In the left panel of Fig~\ref{f:hebump} we show the run of [Na/Fe] for
both clusters; to take into account the different absolute magnitudes of the
bumps in the two GCs because of the different metallicities we normalize
the $M_V$ of the individual stars to the bump ones (+0.41 and +0.58 in NGC~6752 and
NGC~6218, respectively). In the right panel we
plot the average [Na/Fe] ratios as a function of absolute magnitude.
The results are in fair agreement with the theoretical predictions,
taking also into account the fact that
the inclusion of NGC~6752 
should somewhat smear the effect, since the maximum "signal" in this
kind of plot is obtained when the balance of the two different components is
about 50-50\%. In NGC~6752 a large fraction of stars seems to be part of the
O-poor component, as evident from Figure~\ref{f:histom62m67} and from the
fact that the majority of unevolved stars are found to be very N-enriched by
Carretta et al. (2005). On the other hand, NGC~6218 would be better suited for
testing the prediction by Salaris et al. (2006), since in this case we are
comparing almost ``pure" populations, equally divided in O-poor and O-rich
components, but we had to compromise between enhanced signal and 
smaller numbers.

\begin{figure*}
\centering
\includegraphics[bb=30 430 570 700, clip, scale=.8]{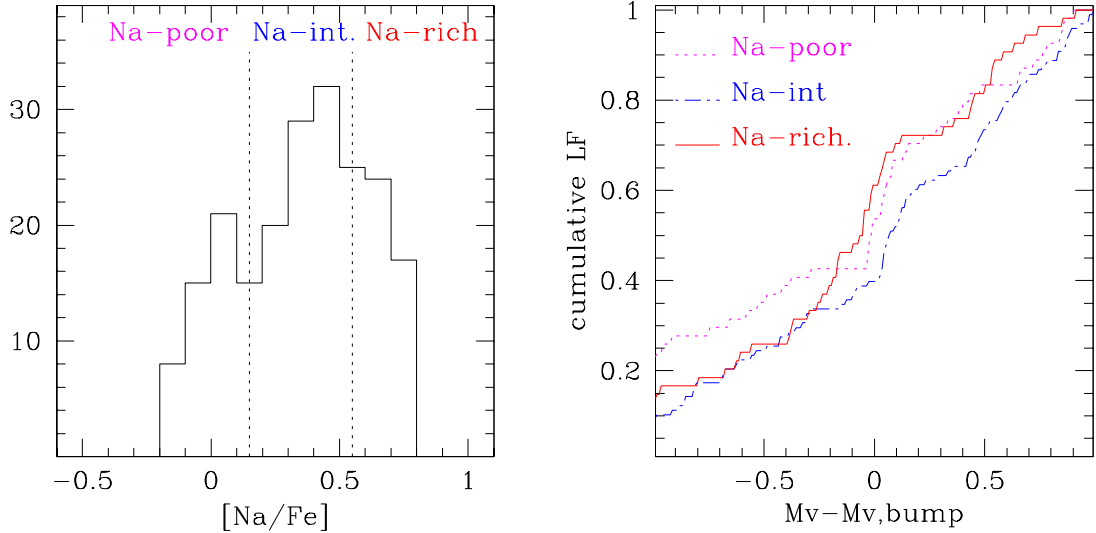}
\caption{Left panel: [Na/Fe] distribution in NGC~6218 and NGC~6752, after correction
for the small trend with magnitude and definition of the Na-poor, intermediate
and rich samples. Right panel: cumulative LFs of the three samples, normalized
to each sample.}
\label{f:flc}
\end{figure*}

We decided to try to have a direct look at the LFs of the
Na-rich and Na-poor populations. We corrected our [Na/Fe] abundances
for the small trend with temperature to eliminate any residual spurious effect.
We divided our stars in three subsamples, indicated in the left
panel of Fig.~\ref{f:flc}: Na-poor, Na-intermediate and Na-rich.
We computed the cumulative LFs (normalized to the number of stars in the
three subsamples) and the part near the RGB-bump is shown in the right
panel of the same figure. It appears that the Na-rich components has a
slightly brighter RGB-bump level, while the Na-poor and Na-intermediate have
fainter, very similar levels. The differences are difficult to appreciate from 
our smallish sample, but we estimate them to be about 0.05 mag between the
Na-rich stars and the others.
From Salaris et al. (2006) we estimate the dependence of the RGB-bump magnitude
from helium abundance to be of about 0.01 mag for each 0.01 variation in Y.
Hence, 0.05 magnitudes between the bumps would translate into an enhancement
of about 0.05 of Y (i.e., Y $\sim$ 0.30 compared to a normal Y = 0.25) for the
Na-rich population, while the Na-intermediate stars do not appear to have
a noticeably different Y.

We conclude that {\em for the first time to our knowledge it is possible to
actually observe the signature of the existence of distinct populations of
stars with different He content in globular clusters only using stars on the
RGB.} That means to be able to independently constrain HB models. 
Admittedly, our evidence is not conclusive, due to the limited
statistics, but the first step in this direction is encouraging. When our
sample of GCs will be entirely analyzed, we will have several other clusters 
with datasets of stars bracketing the RGB-bump and a higher statistical
significance of the plots in Figs~\ref{f:hebump} and \ref{f:flc} will be 
available. Interestingly, Caloi \& D'Antona (2005) interpreted an RGB clump
brighter in M~13 than in M~3 as a possible indication of a different 
(increased) He content in M~13.

\section{Anticorrelation and orbital parameters}

In the previous section we have seen that NGC~6218 is a classical
second-parameter cluster regarding the HB morphology, shows a Na-O anticorrelation
well defined although not extreme, and possibly presents a significant
difference in the average abundance ratios below and above a 
discontinuity at the level of RGB-bump, the likely signature of the presence of
two distinct stellar components formed with different He content.

The only other peculiarity known for this cluster concerns its mass function.
Recently De Marchi et al. (2006, hereinafter DMPP) found that the mass function
(MF) in NGC~6218 is surprisingly flat. The difference 
with most other GCs is that the flattening of the MF
is observed near the cluster half-mass radius, whereas this kind of
flat MF is typically found in the core of GCs and is ascribed to 
relaxation and mass segregation.
By analogy with other clusters showing a flat or dropping MF near the half-mass
radius (namely, NGC~6712 and Pal~5), well known to be subjected to severe tidal
stripping of low mass stars due to repeated interaction with the Galactic
potential well, DPMM suggested that the observed MF in NGC~6218 must be the
result of severe loss of low mass stars in this cluster too.

Our observed sample is located just around and beyond  the half-mass radius of
the cluster and we may try to see if there is a relationship between the
dynamical history of NGC~6218 and the observed chemical anomalies. In order to
answer this question we need to better quantify the relevance and extension of
these anomalies.
Very recently Carretta (2006) proposed the interquartile range (IQR) of
a distribution of abundance ratios as an optimal tool to quantitatively
define the extension of chemical inhomogeneities within a cluster and in
comparison with other clusters. Following his procedure, we computed the
IQR([O/Na])=+0.80 along the Na-O anticorrelation in NGC~6218.

This value is very well correlated with the analogous quantity for the Mg-Al
anticorrelation (see Carretta 2006, figure 11, left panel), hence it is a
quite good quantitative estimate of the amount of chemical variations in a
cluster. The same IQRs can be computed for the [O/Fe] and [Na/Fe] ratios.

\begin{figure}
\centering
\includegraphics[bb=20 170 580 670, clip, scale=0.45]{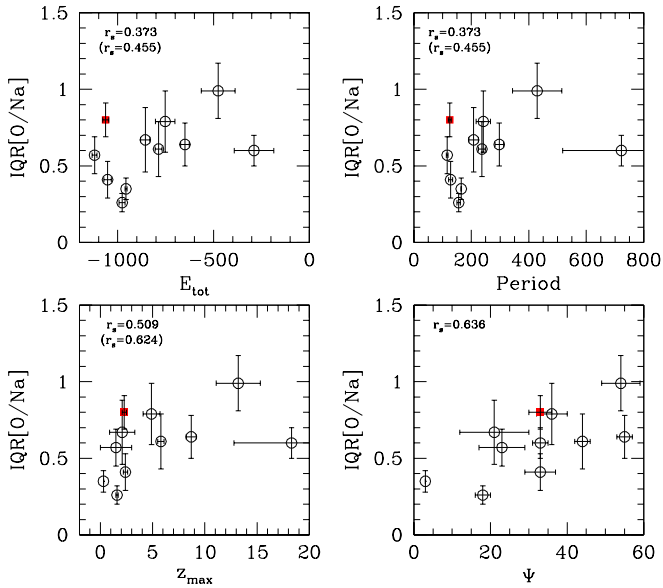}
\caption{Extension of the chemical anomalies in Galactic globular clusters, as
represented by the interquartile range IQR([O/Na]) along the Na-O
anticorrelation, as a function of the orbital parameters: total energy of the
orbit $E_{\rm tot}$ (in units of 10$^2$ km$^2$ s$^{-2}$; top-left panel),
period P (in units of 10$^6$ yrs; top-right panel) of the cluster orbit, 
maximum height above the galactic plane (in kpc; bottom-left panel) and 
orbit inclination (degrees; bottom-right panel). All the orbital parameters are
from Dinescu et al. (1999). The Spearman rank coefficient is shown in each 
panel; values in parenthesis are obtained by excluding M~5 (see Carretta 2006).
The filled square indicates NGC~6218, from the present study. }
\label{f:orb}
\end{figure}

Armed with these useful tools, we explored the relationship between
chemical anomalies and structural parameters, as done by Carretta (2006).
Adding our values for NGC~6218 we confirm that the spread in the various
distributions does not seem to depend directly on the HB type. There seems 
to be a trend
for increasing spread as the cluster mass increases and a more visible
trend for the extension of chemical anomalies to increase with increasing
ellipticity of the cluster.

However, Carretta (2006) discovered that the shape of clusters -represented by
the ellipticity- is strictly related to their orbital features, and in turn this
lead for the first time to the discovery of correlations between chemical
anomalies and orbital parameters in GCs. Our study confirms (see
Fig.~\ref{f:orb}) once more his finding: the IQR([O/Na]) of NGC~6218 
follows the correlations with orbital parameters such as the total energy
$E_{tot}$, the period P of the cluster orbit in the Galaxy, the maximum height
$z_{max}$ above the plane and the inclination angle $\Psi$ with respect to the
Galactic plane. The orbital parameters are taken from Dinescu et al. (1999) for
all clusters, for consistency.

All these relations must be interpreted as a frozen snapshot of the initial
conditions (both in chemistry and in kinematics) existing at the epoch of the
major star formation in the GCs, since the derived orbital characteristics are
the results of averages over several orbital periods. What these correlations
are telling us is simply that clusters (or proto clusters) formed more away
from the Galactic plane and left relatively undisturbed afterwards have more
chances to efficiently retain their gas and to develop a more extended
distribution along the Na-O anticorrelation.

NGC~6218 joins to this pattern. Its currently observed mass is about
one fifth of the original, according to DMPP. This suggests that:
\begin{itemize}
\item[i)] the global anticorrelation must be a phenomenon established early in
the cluster history and widespread over the bulk of cluster stars, since it is
still fully observed even if the majority of the cluster mass has likely been
lost following gravitational shocking and tidal stripping
\item[ii)] the average orbital parameters are very useful to keep tracks
of the original conditions existing at the moment of cluster formation and 
during the early
evolution, insensitive of the changes due to the repeated interactions
with the major sources of gravitational shocking. This likely happens 
because the time interval between the very first stellar generation and the
formation of the second one in a cluster (about 10$^8$ yrs) is of the same
order of magnitude of a single orbital period.
\end{itemize}

\section{Summary}

In this paper we have derived atmospheric parameters and elemental abundances
of Na and O for 79 red giant stars in the globular cluster NGC~6218 observed
with the multifiber spectrograph FLAMES.

Atmospheric parameters for all targets were obtained from the photometry.
From the analysis of the GIRAFFE spectra we derived an average metallicity 
[Fe/H]$=-1.31\pm 0.004\pm 0.028$ dex (random and systematic errors)
(with an rms=0.03 dex, 79 stars), without indication of intrinsic
star-to-star scatter. The errorbars do not include systematic effects 
common to all clusters analyzed in this series of papers.

The [Na/Fe] versus [O/Fe] ratios follow the well known Na-O anticorrelation, as
in all other GCs examined so far, but a few considerations are due.

In particular, the Na and O distribution does not resembles that of NGC~6752,
another intermediate-metallicity cluster with a blue HB morphology: in NGC~6752
the star distribution is skewed toward a predominance of O-poor/Na-rich stars,
whereas in NGC~6218 O-poor and O-rich stars seem to be represented in almost
equal proportion.

We found that the distributions for stars fainter and brighter than the
RGB-bump in NGC~6218 are statistically different, with brighter stars
showing on average a higher Na and a lower O content. Moreover, a step-like
feature in the abundance ratios distribution at the bump level is clearly
present in NGC~6218, and much less evident in NGC~6752.
We interpret this behavior as the expected predictions by Salaris et al.
(2006) that luminosity functions of He-rich and He-poor stars are distinctively
different in the RGB-bump region. Adding together data for [Na/Fe]
ratios in both NGC~6752 and NGC~6218 to gain in statistical significance, we
detected for the first time the different bumps in the LFs of the stellar
components formed with different initial He content in GCs.

Notwithstanding its peculiar mass function and the supposed loss of
about four-fifths of the original mass, NGC~6218 seems to follows the
correlations found by Carretta (2006) between the orbital parameters and
the extension of the chemical anomalies in GCs.

In the future we will continue the analysis of the other clusters in our
sample, progressively adding more and more stars observed in the RGB-bump
region. The goal of obtaining very accurate luminosity levels for the peak of
the distributions of He-rich and He-poor stars in GCs is thus within reach: in
turn, this will allow us to estimate quantitatively the degree of He-enrichment
in cluster stars, a precious constraint to evolutionary stellar models of 
intermediate mass AGB-stars and to the modeling of the birth and early
evolution of the globular clusters.

\begin{acknowledgements}
This
publication makes use of data products from the Two Micron All Sky Survey,
which is a joint project of the University of Massachusetts and the Infrared
Processing and Analysis Center/California Institute of Technology, funded by
the National Aeronautics and Space Administration and the National Science
Foundation. This work was partially funded by the Italian MIUR
under PRIN 2003029437. We also acknowledge partial support from the grant
INAF 2005 ``Experimental nucleosynthesis in clean environments". 
\end{acknowledgements}

\end{document}